\documentclass[11pt]{article}
\usepackage{aip}	
\newcommand{\pard}[2]{\frac{\partial #1}{\partial #2}}  
\def \ccomma{\raise 2pt\hbox{,}} 
\def \D {\hbox{d}}

\begin{document}

\date{19~December~2001}
\title{Integration of a generalized H\'enon-Heiles Hamiltonian}
\author{C.~Verhoeven\dag, M.~Musette\dag, and R.~Conte$^*$\\[6pt]
\dag Dienst Theoretische Natuurkunde, Vrije Universiteit Brussel\\ 
Pleinlaan 2, B--1050 Brussels, Belgium\\ 
E-mail: cverhoev@vub.ac.be and mmusette@vub.ac.be\\[5pt]
* Service de physique de l'\'etat condens\'e, CEA--Saclay\\ 
F--91191 Gif-sur-Yvette Cedex, France\\ 
E-mail:  Conte@drecam.saclay.cea.fr
} 

\maketitle
PACS: 45.20.Jj--Lagrangian and Hamiltonian mechanics; 05.45.Yv--Solitons 

\begin{abstract}
The generalized H\'enon-Heiles Hamiltonian 
$H=1/2(P_X^2+P_Y^2+c_1X^2+c_2Y^2)+aXY^2-bX^3/3$ 
with an additional nonpolynomial 
term $\mu Y^{-2}$ is known to be Liouville integrable 
for three sets of values of $(b/a,c_1,c_2)$. 
It has been previously integrated by  
genus two theta functions only in one of these cases. 
Defining the separating variables of the Hamilton-Jacobi equations, 
we succeed here, in the two other cases, to integrate the equations of
motion with hyperelliptic functions. 
\end{abstract}

\section{Introduction}

The generalization of the H\'enon-Heiles Hamiltonian defined by 
\begin{eqnarray}
\hspace{-15pt}
\label{HH:gen}
H & \equiv &
  \frac{1}{2}(P_X^2+P_Y^2+c_1X^2+c_2Y^2)+aXY^2-\frac{b}{3}X^3
 +\frac{1}{32a^2}\frac{\mu}{Y^2}\ccomma\
a\not=0, 
\\
\hspace{-15pt}
X' & = & P_X,\
Y'=P_Y,\\
\hspace{-15pt}
X''& = & -aY^2+bX^2-c_1 X,\\
\hspace{-15pt}
Y''& = & -2aXY-c_2Y+\frac{1}{16a^2}\frac{\mu}{Y^3}\ccomma\
\mu \hbox{ arbitrary,}
\end{eqnarray}
is integrable in the sense of Liouville in three cases 
\cite{BSV:1982,CTW:1982,GDP:1982}, namely 
\begin{eqnarray}
&&\frac{b}{a}=-1,\quad c_1=c_2,\\ 
&&\frac{b}{a}=-6,\quad c_1,c_2\hbox{ arbitrary,}\\
&&\frac{b}{a}=-16,\quad c_1=16c_2,
\end{eqnarray}
and is equivalent to the reduction $\xi=x-ct$ of three fifth order
soliton equations
\cite{For:1991}, respectively the
Sawada-Kotera (SK), KdV$_5$ and Kaup-Kupershmidt (KK)
equations, by applying the translation
{\setlength \arraycolsep{2pt}\begin{eqnarray}
\hbox{SK: }     &&u=X+\frac{c_2}{2a}\ccomma\qquad
 c=c_1c_2,\\
\hbox{KdV$_5$: }&&u=X+\frac{c_1+4c_2}{20a}\ccomma\quad
 c=\frac{1}{10}(-3c_1^2-16c_1c_2+48c_2^2),\\
\hbox{KK: }     &&u=X+\frac{c_2}{2a}\ccomma\qquad
 c=c_1c_2.
\end{eqnarray}}
The reduced partial differential equations (PDE's) are respectively
{\setlength \arraycolsep{1pt}\begin{eqnarray}
\label{HH:SK4}  \hbox{SK: }     &&
u^{(4)}+10auu''+\frac{20}{3}a^2u^3-cu+4aE-\frac{c_1c_2^2}{3a}=0,\\[5pt]
\label{HH:KdV54}\hbox{KdV$_5$: }&&
u^{(4)}+20auu''+40a^2u^3+10au'^2-cu\nonumber\\
&&+4aE+\frac{1}{100a}(c_1+4c_2)(c_1^2-12c_1c_2+16c_2^2)=0,\\[5pt]
\label{HH:KK4}  \hbox{KK: }     &&
u^{(4)}+40auu''+\frac{320}{3}a^2u^3+30au'^2-cu+4aE-\frac{c_1c_2^2}{3a}=0,
\end{eqnarray}} 
in which $E$ is the constant value of the Hamiltonian $H$. 
Therefore we will further use these names for refering to the respective 
integrable cases of the Hamiltonian.

The problem which we address is to find the separating variables and to
explicitly integrate.

The general solution for the KdV$_5$ case has been obtained in terms of 
hyperelliptic functions \cite{Dra:1919,AnP:1983,Woj:1984} by separation
of the 
variables of 
the Hamilton-Jacobi
equation in parabolic  coordinates.

For $\mu=0$, the equations of motion for the SK and KK cases have been 
integrated \cite{Cha:1911,AiS:1972,RGC:1993} in terms of
elliptic functions.

In this paper we give the general solution of the equations of motion
for the SK 
and KK cases in terms 
of hyperelliptic functions. This is achieved by separation of the
variables of 
the Hamilton-Jacobi 
equation, where the canonical transformation between SK and KK
\cite{Bak:1995} 
plays an important role.

In section \ref{section2},
we consider the SK case with $\mu=0$ as a starting point to treat  the
general 
case ($\mu$ arbitrary). 
In  section \ref{section3}, 
we recover the canonical transformation between SK and KK,
starting from the factorization of the scattering operator   associated
with these equations and
using the link with the Fordy-Gibbons equation \cite{FoG:1980}. 
In section \ref{section4}, 
we give the general solution for KK, using the canonical transformation 
previously  obtained. 
In section \ref{section5}, 
we use the
fact that the canonical transformation is invertible for obtaining the 
general solution of the SK case. 
In both cases, in the limit $\mu\rightarrow 0$, 
we recover the previous results \cite{AiS:1972,RGC:1993}.
In section \ref{section6}, 
we compare our method of integration, starting from the Hamiltonian
system, with the method used by Cosgrove
\cite{Cos:2000} for integrating the fourth order ODE's (\ref{HH:SK4})
and (\ref{HH:KK4}).

\section{Particular case: Sawada-Kotera for $\mu=0$}
\label{section2}

This is the simplest case and it will be our starting point to define
the separation of variables of the Hamilton-Jacobi equation for KK.

Consider the Hamiltonian in the SK case derived from (\ref{HH:gen}) by setting
$a=\textstyle{\frac{1}{2}}$ and defining $,U=X+c_2,V=Y,c=c_1c_2$
\begin{equation}\label{ham:SK}
H\equiv 
K_{1,0}=\frac{1}{2}(P_U^2+P_V^2)+\frac{1}{2}UV^2+\frac{1}{6}U^3-\frac{c}{2}U,
\end{equation}
which is a constant of motion of the equations
\begin{equation}
\begin{array}{ll}
U''=-\frac{1}{2}(V^2+U^2)+\displaystyle{\frac{c}{2}}\ccomma\\
V''=-UV,
\end{array}
\end{equation} 
where $U,V$ and the derivatives $U'=P_U,V'=P_V$ are functions of the 
independent variable $\xi$ ($'\equiv \frac{\D}{\D\xi}$). 

This system possesses a second constant of motion \cite{AiS:1972}
\begin{equation}
K_{2,0}=-2P_UP_V-U^2V-\frac{V^3}{3}+c V,
\end{equation}
which is in involution with the Hamiltonian, i.e. $\{K_{1,0},K_{2,0}\}=0$.

The separation of variables is defined as
\begin{equation}\label{HHSK:sepv}
\begin{array}{ll}
Q_1=U+V,                  &Q_2=U-V,\\
P_1=\frac{1}{2}(P_U+P_V), &P_2=\frac{1}{2}(P_U-P_V),
\end{array}
\end{equation}
and in those new variables the expressions of $K_{1,0}$ and $K_{2,0}$ are
{\setlength\arraycolsep{2pt}\begin{eqnarray}
\label{H0SK:sep}
K_{1,0}&=&P_1^2+P_2^2+\frac{1}{12}(Q_1^3+Q_2^3)-\frac{c}{4}(Q_1+Q_2),\\
\label{K0SK:sep}
K_{2,0}&=&2(P_2^2-P_1^2)+\frac{1}{6}(Q_2^3-Q_1^3)-\frac{c}{2}(Q_
2-Q_1),
\end{eqnarray}}
such that the equations of motion become
\begin{equation}
\left\{\begin{array}{ll}
Q_1'^2=-\frac{1}{3}Q_1^3+cQ_1-K_{2,0}+2K_{1,0},\\
Q_2'^2=-\frac{1}{3}Q_2^3+cQ_2+K_{2,0}+2K_{1,0}.
\end{array}
\right.
\end{equation}
They possess the general solution
\begin{equation}
\left\{\begin{array}{ll}
Q_1=-12\wp\left(\xi-\xi_1,\frac{c}{12},-\frac{1}{144}(2K_{1,0}-K_{2,0})\right)
\equiv -12\wp_1(\xi),\\
Q_2=-12\wp\left(\xi-\xi_2,\frac{c}{12},-\frac{1}{144}(2K_{1,0}+K_{2,0})\right)
\equiv -12\wp_2(\xi),
\end{array}
\right.
\end{equation}
in terms of the Weierstrass elliptic function $\wp(x-x_0,g_2,g_3)$, 
solution of the first order differential equation
\begin{equation}
\wp'^2=4\wp^3-g_2\wp-g_3,
\end{equation}
such that the solutions become
\begin{equation}\label{HH:sSK0}
\left\{\begin{array}{ll}
U=-6\left(\wp_1(\xi)+\wp_2(\xi)\right),\\ 
V=-6\left(\wp_1(\xi)-\wp_2(\xi)\right).   
\end{array}
\right.
\end{equation}

\section{Canonical transformation between SK and KK.}
\label{section3}

Let us consider, for $\mu$ arbitrary, the equations and the constants of
motion 
associated with the SK and KK Hamiltonians
\cite{GDR:1984,Hie:1987} {\setlength \arraycolsep{2pt}
\begin{eqnarray}
\label{HH:e1SK}\hbox{SK: }&&U'=P_U,\qquad V'=P_V,\\
\label{HH:e2SK}&&U''=-\frac{1}{2}(V^2+U^2)-\frac{c}{2}\ccomma\qquad 
V''=-UV+\frac{\mu}{4V^3}\ccomma\\
\label{HH:HSK}&&K_1=\frac{1}{2}(P_U^2+P_V^2)
+\frac{1}{2}UV^2+\frac{1}{6}U^3-\frac{c}{2}U+
\frac{\mu}{8V^2}\ccomma\\
\label{HH:KSK}&&K_2^2=K_{2,0}^2+\frac{2}{3}\mu U+\mu\frac{P_U^2}{V^2}\ccomma
\\[8pt]
\hbox{KK: }\label{HH:paKK}&&a=\frac{1}{4},\quad c_1=16c_2,\quad 
c=c_1c_2,\quad u=X+2c_2,\quad v=Y,\\
\label{HH:e1KK}&&u'=p_u,\qquad v'=p_v,\\
\label{HH:e2KK}&&
u''=-\frac{1}{4}v^2-4u^2+c,\qquad 
v''=-\frac{1}{2}uv+\frac{\mu}{v^3}\ccomma\\
\label{HH:HKK}&&
k_1=\frac{1}{2}(p_u^2+p_v^2)+\frac{1}{4}uv^2+\frac{4}{3}u^3-cu+
\frac{1}{2}\frac{\mu}{v^2}\ccomma\\
\label{HH:K0KK}&&
k_{2,0}^2=p_v^4-\frac{1}{72}v^6-\frac{1}{12}u^2v^4+up_v^2v^2
-\frac{1}{3}p_up_vv^3+\frac{c}{12}v^4,\\
\label{HH:KKK}&&
k_2^2=k_{2,0}^2+\frac{\mu}{3}u+2\mu\frac{p_v^2}{v^2}+\frac{\mu^2}{v^4}\cdot
\end{eqnarray}}
The reason why the expressions $k_{2,0}$ and $k_2$ are defined by their square
will appear soon. 

The two nonlinear partial differential equations SK and KK
{\setlength\arraycolsep{1pt} \begin{eqnarray}
\hbox{SK: }&&U_t+(U_{xxxx}+5UU_{xx}+\frac{5}{3}U^3)_x=0,\\
\hbox{KK: }&&u_t+(u_{xxxx}+10uu_{xx}+\frac{20}{3}u^3+30u_x^2)_x=0,
\end{eqnarray}}
whose reductions $\xi=x-ct$ are
(\ref{HH:SK4}) and (\ref{HH:KK4}), 
respectively obtained from the systems (\ref{HH:e2SK})--(\ref{HH:HSK}) 
and (\ref{HH:e2KK})--(\ref{HH:HKK}) 
by elimination of the variables $V$ and $v$, 
possess a Lax pair with a third  order scattering problem
($L\psi=\lambda \psi$)
\cite{SaK:1977,Kau:1980}. 
The scattering operators can be factorized in the following way
{\setlength\arraycolsep{2pt}\begin{eqnarray}
\hbox{SK: }&&L\equiv 
\partial_x^3+U\partial_x=(\partial_x-w)(\partial_x+w)\partial_x,\\
\hbox{KK: }&&L\equiv 
\partial_x^3+2u\partial_x+u_x=(\partial_x+w)(\partial_x)(\partial_x-w),
\end{eqnarray}}
such that the solutions of the PDE's are related through a B\"acklund 
transformation
{\setlength \arraycolsep{2pt}\begin{eqnarray}
\hspace{-125pt}\label{HH:MiSK}\hbox{SK: }&&U=w_x-w^2,\\
\hspace{-125pt}\label{HH:MiKK}\hbox{KK: }&&u=-w_x-\frac{1}{2}w^2.
\end{eqnarray}}
with the solution $w$ of the Fordy Gibbons equation 
\cite{FoG:1980}
\begin{equation}
w_t+(w_{4x}-5w_xw_{xx}-5w^2w_{xx}-5ww_x^2+w^5)_x=0.
\end{equation}

The reduction $\xi =x-ct$ of this equation can be solved for $w(\xi)$, 
either by eliminating $U$ and $w'$ between (\ref{HH:MiSK}) and the
equations 
of motion (\ref{HH:e2SK})
\begin{equation}
w=\frac{1}{2}\frac{\sqrt{-\mu}}{V^2}-\frac{V'}{V}\ccomma
\end{equation}
or by eliminating $u$ and $w'$ between (\ref{HH:MiKK}) and the equations
of 
motion (\ref{HH:e2KK})
\begin{equation}
w=2\frac{\sqrt{-\mu}}{v^2}+2\frac{v'}{v}\ccomma
\end{equation}
such that, defining $\lambda^2=-\mu$ and
{\setlength\arraycolsep{2pt}
\begin{eqnarray}
\Gamma&=&6(VK_{2,0}+\lambda P_U),\\[6pt]
\Omega&=&
48(3v^4k_{2,0}^2+6\lambda uv^5p_v+12\lambda p_v^3v^3-\lambda v^6p_u
\nonumber\\
&&
+3\lambda ^2uv^4+18\lambda ^2v^2p_v^2+12\lambda^3vp_v+3\lambda^4),
\end{eqnarray}}
the canonical transformation is given by \cite{Bak:1995,BRW:1994}
{\setlength\arraycolsep{2pt}
\begin{eqnarray}
\label{HHu:KKSK}
u&=&-\frac{3}{2}\left(-\frac{P_V}{V}+\frac{\lambda}{2V^2}\right)^2-U,\qquad 
v^2=\frac{\Gamma}{V^2}\ccomma
\\[6pt]
p_u&=&\frac{1}{V^3}(3P_V^3+3UV^2P_V-P_UV^3)\nonumber\\
&&-\frac{3\lambda}{2V^6}
\left(UV^4+3V^2P_V^2-\frac{3}{2}\lambda VP_V+\frac{\lambda^2}{4}\right), 
\nonumber
\\[6pt]
\label{HHp:KKSK}
p_v&=&\frac{1}{4V^2}\left(-2P_V+\frac{\lambda}{V}\right)\sqrt{\Gamma}-\lambda
\frac{V}{\sqrt{\Gamma}}\ccomma\\[10pt]
\label{HHU:SKKK}
U&=&-6\left(\frac{p_v}{v}+\frac{\lambda}{v^2}\right)^2-u,\qquad 
V^2=\frac{\Omega}{4v^8}\ccomma\\[6pt]
P_U&=&\frac{1}{v^3}(12p_v^3+6uv^2p_v-v^3p_u)\nonumber\\
&&+\frac{3\lambda}{v^6}(2uv^4+12v^2p_v^2+12\lambda 
vp_v+4\lambda^2),\nonumber\\[6pt]
\label{HHP:SKKK}P_V&=&-\frac{1}{v^5}\left(p_v+\frac{\lambda}{v}\right)
\sqrt{\Omega}+\lambda\frac{v^4}{\sqrt{\Omega}}\cdot
\end{eqnarray}}

\section{General solution of the Kaup-Kupershmidt case}
\label{section4}

Starting from the separation of variables (\ref{HHSK:sepv}) 
and the canonical transformation (\ref{HHU:SKKK})--(\ref{HHP:SKKK})
in the case $\mu=0$,
we consider the transformation defined by 
\cite{RGC:1993} on the basis of Painlev\'e analysis
\begin{equation}\label{HHKK:sepv}
\left\{\begin{array}{ll}
q_1=\displaystyle{-6\frac{p_v^2-k_{2,0}}{v^2}-u},\\[6pt]
q_2=\displaystyle{-6\frac{p_v^2+k_{2,0}}{v^2}-u},\\[6pt]
p_1=\displaystyle{\frac{1}{2v^3}(12p_v^3+6uv^2p_v-v^3p_u-12p_vk_{2,0})},\\[6pt]
p_2=\displaystyle{\frac{1}{2v^3}(12p_v^3+6uv^2p_v-v^3p_u+12p_vk_{2,0})}.\\[6pt]
\end{array}
\right.
\end{equation} 
This inverts to
\begin{equation}
\left\{\begin{array}{ll}
u=\displaystyle{-6\left(\frac{p_2-p_1}{q_2-q_1}\right)^2-\frac{1}{2}(q_1+q_2)},
\\[6pt]
v^2=\displaystyle{\frac{12k_{2,0}}{q_1-q_2}}\ccomma\\[6pt]
p_u=\displaystyle{24\left(
\frac{p_1-p_2}{q_1-q_2}\right)^3+2(p_1-p_2)\frac{q_1+q_2}{q_1-q_2}
+2\frac{p_1q_2-p_2q_1}{q_1-q_2}}\ccomma\\[6pt]
p_v^2=\displaystyle{12k_{2,0}\frac{(p_2-p_1)^2}{(q_1-q_2)^3}}\cdot
\end{array}
\right.
\end{equation}

Taking account that $k_{2,0}$ is no more a constant of motion, 
this change of variables will appear be useful to find the 
general solution for KK. 
In those new variables, the Hamiltonian system (\ref{HH:HKK}) becomes 
{\setlength\arraycolsep{2pt}\begin{eqnarray}
\label{HHKK:Hsep}
H&\equiv &k_1=
p_1^2+p_2^2+\frac{1}{12}(q_1^3+q_2^3)-\frac{c}{4}(q_1+q_2)+
\frac{\mu}{24}\frac{q_1-q_2}{k_{2,0}}\ccomma\\[6pt]
\label{HHKK:k0sep}
k_{2,0}&=&2(p_2^2-p_1^2)+\frac{1}{6}(q_2^3-q_1^3)-\frac{c}{2}(q_2-q_1),\\[6pt]
\label{HHKK:e1sep}
q_1'&=&2p_1+\frac{\mu}{6}\frac{(q_1-q_2)p_1}{k_{2,0}^2}\ccomma\\[6pt]
\label{HHKK:e2sep}
q_2'&=&2p_2-\frac{\mu}{6}\frac{(q_1-q_2)p_2}{k_{2,0}^2}\cdot
\end{eqnarray}} 
Therefore, defining $f(q_i,p_i)\equiv
p_i^2+\textstyle{\frac{1}{12}}q_i^3-\textstyle{\frac{c}{4}}q_i\quad
(i=1,2)$, 
the Hamilton-Jacobi equation is separated 
\begin{eqnarray}
& &
k_1\Big(f(q_1,p_1)-f(q_2,p_2)\Big)=f^2(q_1,p_1)-f^2(q_2,p_2)+
\frac{\mu}{48}(q_1-q_2),
\\
& & p_i=\pard{S}{q_i}\quad(i=1,2)\cdot
\nonumber
\end{eqnarray}

We can express the second constant of motion $k_2^2$ in two equivalent ways
\begin{eqnarray}
\label{HHKK:ksep1}
&&k_2^2=-\frac{\mu}{3}q_1+(k_{2,0}+\frac{\mu}{12}\frac{q_1-q_2
}{k_{2,0}})^2,\\
\label{HHKK:ksep2}\hbox{or}&&k_2^2=-\frac{\mu}{3}q_2+(k_{2,0}-\frac{\mu}{12}
\frac{q_1-q_2}{k_{2,0}})^2,
\end{eqnarray}
so that the elimination of 
$\mu (q_1-q_2)/k_{2,0}$
between
(\ref{HHKK:Hsep}), (\ref{HHKK:ksep1}) and 
(\ref{HHKK:Hsep}), (\ref{HHKK:ksep2}) yields
\begin{eqnarray}
\label{HHKK:ksep3}
&&k_2^2=-\frac{\mu}{3}q_1+(-4p_1^2-\frac{q_1^3}{3}+cq_1+2k_1)^
2,\\
\label{HHKK:ksep4}
\hbox{or}&&k_2^2=-\frac{\mu}{3}q_2+(4p_2^2+\frac{q_2^3}{3}-cq_
2-2k_1)^2.
\end{eqnarray}
The elimination of $p_1$ between (\ref{HHKK:ksep3}) and (\ref{HHKK:e1sep}),
and             of $p_2$ between (\ref{HHKK:ksep4}) and
(\ref{HHKK:e2sep}) 
yields                                                                  
{\small
\begin{eqnarray}
\label{HHKK:e3sep}&&\hspace{-30pt}
q_1'=\sqrt{2k_1-\frac{q_1^3}{3}+cq_1-\sqrt{k_2^2+\frac{\mu}{3}q_1}}
\left(1+\frac{\mu}{3}\frac{q_1-q_2}{\left(\sqrt{k_2^2+\frac{\mu}{3}q_2}+
\sqrt{k_2^2+\frac{\mu}{3}q_1}\right)^2}\right)\ccomma\\
\label{HHKK:e4sep}&&\hspace{-30pt}
q_2'=\sqrt{2k_1-\frac{q_2^3}{3}+cq_2+\sqrt{k_2^2+\frac{\mu}{3}q_2}}\left(1-
\frac{\mu}{3}\frac{q_1-q_2}{\left(\sqrt{k_2^2+\frac{\mu}{3}q_2}+
\sqrt{k_2^2+\frac{\mu}{3}q_1}\right)^2}\right)\cdot
\end{eqnarray}}
In the case $\mu=0$, 
the differential equations for $q_1$ and $q_2$ are separated and their 
solution is expressed in terms of the Weierstrass elliptic function
\begin{eqnarray}
&&q_{1,0}=
-12\wp\left(\xi-\xi_1,\frac{c}{12},-\frac{1}{144}(2k_{1,0}-k_{2,0})\right)
\equiv -12\wp_1(\xi),\\
&&q_{2,0}=
-12\wp\left(\xi-\xi_2,\frac{c}{12},-\frac{1}{144}(2k_{1,0}+k_{2,0})\right)
\equiv -12\wp_2(\xi),
\end{eqnarray}
so that the solution for (\ref{HH:e2KK}) in the case $\mu=0$ is 
\cite{RGC:1993}
\begin{eqnarray}
\label{HH:sKK0u}&&
u=-\frac{3}{2}
\left(\frac{\wp_1'(\xi)-\wp_2'(\xi)}{\wp_1(\xi)-\wp_2(\xi)}\right)^2
+6(\wp_1(\xi)+\wp_2(\xi)),\\
\label{HH:sKK0v}&&
v^2=\displaystyle{\frac{k_{2,0}}{\wp_2(\xi)-\wp_1(\xi)}}\cdot
\end{eqnarray}
In the case $\mu\neq 0$, let us
introduce the new variables
\begin{equation}
s_1=\sqrt{3\frac{k_2^2}{\mu}+q_1},\qquad s_2=-\sqrt{3\frac{k_2^2}{\mu}+q_2},
\end{equation}
which transform the equations (\ref{HHKK:e3sep}) and (\ref{HHKK:e4sep}) 
into
\begin{eqnarray}
\label{HH:hype1}&&\hspace{-10pt}
s_1'=\sqrt{2k_1-\frac{1}{3}(s_1^2-3\frac{k_2^2}{\mu})^3
               +c(s_1^2-3\frac{k_2^2}{\mu})-\sqrt{\frac{\mu}{3}}s_1}
\left(\frac{1}{s_1-s_2}\right)\ccomma\\
\label{HH:hype2}&&\hspace{-10pt}
s_2'=-\sqrt{2k_1-\frac{1}{3}(s_2^2-3\frac{k_2^2}{\mu})^3
               +c(s_2^2-3\frac{k_2^2}{\mu})-\sqrt{\frac{\mu}{3}}s_2}
\left(\frac{1}{s_1-s_2}\right)\cdot
\end{eqnarray}

Defining
\begin{equation}\label{HH:hypc}
P(s)=2k_1-\frac{1}{3}\left(s^2-3\frac{k_2^2}{\mu}\right)^3
    +c\left(s^2-3\frac{k_2^2}{\mu}\right)-\sqrt{\frac{\mu}{3}}s,
\end{equation}
the system (\ref{HH:hype1})--(\ref{HH:hype2}) can be solved by inversion
of the hyperelliptic integrals  
\begin{eqnarray}
\label{hyp1:HH}&&\int_{\infty}^{s_1}{\frac{\hbox{d}s}{\sqrt{P(s)}}}+
\int_{\infty}^{s_2}{\frac{\hbox{d}s}{\sqrt{P(s)}}}=k_3,\\
\label{hyp2:HH}&&\int_{\infty}^{s_1}{\frac{s\hbox{d}s}{\sqrt{P(s)}}}+
\int_{\infty}^{s_2}{\frac{s\hbox{d}s}{\sqrt{P(s)}}}=\xi+k_4,
\end{eqnarray}
which define $s_1$ and $s_2$ as multivalued 
functions of
$\xi$ \cite{Kow:1889,Gol:1953}.

The general solution of the equations of motion (\ref{HH:e2KK}) 
in the case $\mu\neq 0$ is
\begin{eqnarray}
\label{HH:sKKu}&&
u=-\frac{1}{2}(s_1^2+s_2^2)+\frac{3}{\mu}k_2^2
  -\frac{3}{2}\left(\frac{s_1'+s_2'}{s_1+s_2}\right)^2\ccomma\\
\label{HH:sKKv}&&
v^2=\frac{2\sqrt{3\mu}}{s_1+s_2}\cdot
\end{eqnarray}
As they are rational symmetric combinations of $s_1$ and $s_2$ and their
derivatives, $u$ and $v^2$ are single-valued functions of $\xi$.

In the variables $q_1,q_2$ this solution is expressed as 
{\setlength\arraycolsep{1pt}\begin{eqnarray}
u&=&-\frac{3}{2}\left(\frac{\sqrt{2k_1-\frac{q_1^3}{3}+cq_1-\sqrt{k_2^2+
\frac{\mu}{3}q_1}}-
\sqrt{2k_1-\frac{q_2^3}{3}+cq_2
+\sqrt{k_2^2+\frac{\mu}{3}q_2}}}{q_1-q_2}\right)^
2\nonumber\\
&&-\frac{1}{2}(q_1+q_2),\\
v^2&=&
 6\frac{\sqrt{k_2^2+\frac{\mu}{3}q_1}+\sqrt{k_2^2+\frac{\mu}{3}q_2}}{q_1-q_2}
\ccomma
\end{eqnarray}}
which clearly goes to (\ref{HH:sKK0u})--(\ref{HH:sKK0v}) 
in the limit $\mu\rightarrow 0$.

\section{General solution of the Sawada-Kotera case}
\label{section5}

We start from the general solution (\ref{HH:sKKu}), (\ref{HH:sKKv}) for KK
and apply the canonical transformation (\ref{HHU:SKKK})
to obtain the general solution for the SK Hamiltonian system

{\setlength \arraycolsep{2pt}\begin{eqnarray}
\label{HH:sSKU}\hspace{-10pt}U&=&\sqrt{-3}(s_1'+s_2')+s_1^2+s_1s_2+s_2^2-
\frac{3}{\mu}K_2^2,\\
\label{HH:sSKV}\hspace{-10pt}V^2&=&-2\sqrt{-3}(s_1+s_2)(s_1s_1'+s_2s_2')
+2(s_1+s_2)^2\left(s_1^2+s_2^2-\frac{9K_2^2}{2\mu}\right)\cdot 
\end{eqnarray}}

Let us check that the limit of this solution when $\mu\rightarrow 0$ is 
(\ref{HH:sSK0}).
The expression (\ref{HH:sSKU}) for $U$ can also be written as
{\setlength\arraycolsep{2pt} 
\begin{eqnarray}
U&=&\frac{1}{2}(s_1^2+s_2^2)-\frac{3}{\mu}K_2^2
+\frac{3}{2}\left(\frac{s_1'+s_2'}{s_1+s_2}\right)^2\nonumber\\
&&-\frac{3}{2}\left(\frac{s_1'+s_2'}{s_1+s_2}
                    +\sqrt{-\frac{1}{3}}(s_1+s_2)\right)^2.
\end{eqnarray}
}
Since in the limit $\mu\rightarrow 0$ 
\begin{eqnarray*}
&&s_1+s_2={\cal O}(\sqrt{\mu}),\\
&&\frac{s_1'+s_2'}{s_1+s_2}
\rightarrow 
\frac{\wp_1'(\xi)-\wp_2'(\xi)}{\wp_1(\xi)-\wp_2(\xi)}\ccomma
\end{eqnarray*}
one has
\begin{equation}
\lim_{\mu \to 0} U=\frac{1}{2}(Q_1+Q_2). 
\end{equation}

Next, for $V^2$, the expansions 
\begin{displaymath}
2(s_1+s_2)^2\left(s_1^2+s_2^2-\frac{9K_2^2}{2\mu}\right)
=\frac{1}{4}(Q_1-Q_2)^2+{\cal O}(\mu),\ \mu \to 0, 
\end{displaymath}
\begin{displaymath}
(s_1+s_2)(s_1s_1'+s_2s_2')={\cal O}(\sqrt{\mu}),\ \mu \to 0,
\end{displaymath}
provide the limit of $V^2$ in (\ref{HH:sSKV}).
Therefore,
\begin{equation}
\lim_{\mu \to 0} V^2=\frac{1}{4}(Q_1-Q_2)^2.
\end{equation}

\section{Comparison with the results of Cosgrove}
\label{section6}

C.~Cosgrove \cite{Cos:2000} recently integrated 
the ODE's (\ref{HH:SK4}) and (\ref{HH:KK4}) 
with hyperelliptic functions, using the postmultiplier method. 

To compare the two different ways of integration, 
let us recall the vocabulary introduced in Painlev\'e analysis of nonlinear
differential equations, 
making the distinction between fixed and movable constants. 
A constant is called 
\emph{fixed} if it appears explicitly in the differential equation, 
while it is \emph{movable} if it is a constant of integration and therefore
depends on the initial data.

In the Hamiltonian formalism, described by the system 
(\ref{HH:paKK})--(\ref{HH:KKK}) and (\ref{HH:e1SK})--(\ref{HH:KSK}),
$k_1$ (resp.~$K_1$) and $k_2$ (resp.~$K_2$) are movable constants, 
while $\mu$ is fixed (it appears in the equations of motion).

In Cosgrove's paper, 
the first integrals of the fourth order equation he obtained in 
formulas (4.3)--(4.4) and (5.6)--(5.7), 
which are  therefore movable constants, 
correspond in the Hamiltonian formalism to $k_2$ (resp.~$K_2$) and
$\mu$, 
respectively introduced as movable and fixed constants. 
In order to integrate the resulting second order sixth degree
differential 
equation and transform it in a coupled system of first order equations 
for applying the postmultiplier method, 
Cosgrove defined ``suitable''
auxiliary variables chosen in a ``subjective'' way.
Here the link between the canonical variables 
and its expressions (4.5)--(4.6) and (5.3)--(5.4) can be clearly
established. 
Those expressions, which are ``hidden'' variables for the fourth order
differential equation, are nothing else than the canonical variables $v^2$ 
(resp. $V^2$) and $p_v^2$ (resp. $P_V^2$) explicitly defined in the 
Hamiltonian formalism.

\section{Conclusion}

We have proven that the three integrable cases (SK, KdV$_5$, KK) of the 
H\'enon-Heiles Hamiltonian can be integrated in terms 
of hyperelliptic functions. 

We will take advantage of the method developed here to integrate other
Hamiltonian systems with two degrees of freedom and additional 
nonpolynomial terms. 

\section{Acknowledgements}

The authors thank Vadim Kuznetsov and Pol Vanhaecke for interesting and 
useful discussions. 
They acknowledge the financial support of the Tournesol grant T99/040.
MM and CV thank the Belgian government for the financial support
extended 
within the framework of the IUAP contract P4/08.
CV is a research assistant of the Fund for Scientific Research, Flanders.

\end{document}